\documentclass[aps,twocolumn,showpacs,nofootinbib]{revtex4}
\usepackage{graphicx}
\newcommand{\beq}{\begin{equation}}
\newcommand{\eeq}{\end{equation}}
\newcommand{\beqa}{\begin{eqnarray}}
\newcommand{\eeqa}{\end{eqnarray}}

\def\half{\frac{1}{2}}
\def\phi{\varphi}

\def\opone{\leavevmode\hbox{\small1\normalsize\kern-.33em1}}
\def\ket#1{|#1\rangle}

\def\braket#1#2{\langle\, #1\,|\,#2\,\rangle}
\def\binomial#1#2{\left(\begin{array}{c} #1 \\ #2 \\\end{array}\right)}
\def\mean#1{\langle#1\rangle}

\begin{document}

\title{Quantum measurement of spins and magnets, and the classical limit of PR-boxes}
\author{Nicolas Gisin \\
\it \small   Group of Applied Physics, University of Geneva, 1211 Geneva 4,    Switzerland}

\date{\small \today}

\begin{abstract}
When Alice measures all her spin-$\half$ of a large ensemble of $N$ singlets, all along the same direction $\vec a$, she prepares at a distance an ensemble of spins for Bob which, because of statistical fluctuations, have a magnetic moment of the order $\sqrt{N}$. By making $N$ large enough, this magnetic moment can be made arbitrarily large. We show that, nevertheless, Bob can't read out of this large magnetic moment Alice's choice of measurement direction $\vec a$. We also consider stronger than quantum correlations and show that Tsirelson's bound follows from the physical assumption that in the macroscopic limit all measurements are compatible and that this should not lead to signaling. 
\end{abstract}

\maketitle

\section{Introduction}\label{intro}
%====================
The question of how one should apply quantum theory to our macroscopic world, and even the big question whether quantum theory applies at all scales, have been with us since the inception of quantum theory. To illustrate the question, let as study the following little conundrum.

First, consider a single spin $\half$. When measured the result is probabilistic and the quantum state perturbed, except if the spin was in a state without quantum uncertainty. Next, consider a large ensemble of $N$ spins, as in a magnet. Then there is no doubt that one can measure the global magnetization essentially without any disturbance. Now, if Alice and Bob share 2 spins in the singlet state and if Alice measures her spin in a direction we label $z$, then she will get as result $\pm1$ (assuming her measurement is described by the Pauli operator $\sigma_z$). This prepares Bob's spin at a distance in the state $\ket{\mp z}$. Since the mixture of these two states is independent of Alice's measurement direction (and equal to Bob's state obtained by tracing out Alice), there is no signaling from Alice to Bob, as is well known.

But consider now the case where Alice and Bob share a large number $N$ of pairs of spins, each in the singlet state. If Alice measures all of them, individually, in the $z$-direction and adds all her results, then she will get a positive or negative fluctuation around zero of the order $\mu\approx \sqrt{N}$. Because of the quantum correlation, Bob will also get a fluctuation of the order $\mu$, i.e. a magnetization of about $\pm\mu$ in the $z$-direction. By making $N$ large enough, Bob's magnetization $\mu$ can be made arbitrarily large. But then, if the magnetization is arbitrarily large, it may seem that Bob can measure it without significantly perturbing it. Obviously, the same should hold if Alice chooses to measure her spins in the $x$-direction. But then, it seems that Bob could determine the direction in which his magnetization points, either $\pm z$ or $\pm x$. Bob could thus deduce from his magnetization the measurement direction chosen by Alice, without anything carrying this information from Alice to Bob; this would be signaling. Moreover, by enlarging the distance between Alice and Bob and assuming Bob's measurement takes a finite time, this signaling would lead to faster than light communication. But that is impossible. Hence, there must be something wrong in the above story.

In this paper we use apply standard quantum measurement theory, i.e. we couple the spin system under investigation to the pointer of the measuring device, treated quantum mechanically, to show how to resolve this conundrum. We shall see that the size of the system, here the size of the magnetization $\mu$ doesn't suffice to characterize systems that can be measured ``classically''. In our example, the background noise of the randomly oriented spins, although averaging to zero magnetization, can't be ignored. 

Next, we investigate what happens if one replaces the quantum singlet state by stronger than quantum correlations, such as the so called PR-boxes. Following Miguel Navascues \cite{Miguel} and Daniel Rochlich \cite{Daniel}, we argue that any physical box, when there are large ensemble of them, should admit ``classical'' measurements. We show that isotropic noisy PR-boxes \cite{isoPRbox} satisfy this highly plausible physical constrain if and only if the noise is large enough for the correlation to be quantum. We thus recover Tsirelson's bound from a physical assumption, in contrast to previous derivations based on more information theoretical arguments \cite{NoTrivialCommComplexity,NoAdvNLComputation,infoCausality}.

\section{Weak measurements as classical measurements}\label{WM}
%====================================================
The standard description of quantum measurements goes as follows. First, one couples the system to be measured to an auxiliary system called the pointer. The later is initially in the state $\ket{q=0}$, i.e. it point to zero. The coupling between the system and the pointer is assumed so strong and brief that, during that short time, one may safely ignore all hamiltonian evolutions, except the one that describes the coupling \cite{VonNeuman,Aharonov}:
\beq \label{Hc}
H_c=g(t)A\otimes p
\eeq
where $A$ is the operator describing the physical quantity to be measured and $p$ is the translation operator acting on the pointer's position; g(t) is a function with non-zero values only during the short system-pointer interaction time, normalized such that $\int g(t)dt=1$.

Let us illustrate this in the case of $N$ spin $\half$, all in the state $\ket{\vec m}$, with measurement $A=\sum_{j=1}^N \sigma_z^j$, where $\sigma^j$ acts on the $j$-th spin. Assume that the pointer's initial state is $\ket{q=0}=\Phi(x)$, with, for example, the function $\Phi(x)$ a Gaussian:
\beq
\Phi(x)=(2\pi\Delta^2)^{-1/4}exp\{-x^2/4\Delta^2\},
\eeq 
where $\Delta$ is the mean square deviation of the pointer's position. After the interaction (\ref{Hc}) the initial state $\ket{\vec m}^{\otimes N}\otimes\ket{q=0}$ evolves to:
\beq \label{SystPointer}
\Psi_{SP}=\sum_{k=0}^N \braket{k,N}{\vec m^{\otimes N}}\cdot \ket{k,N}\otimes\ket{q=2k-N}
\eeq
where $\ket{k,N}$ is the normalized and symmetrized state of $N$ spins with $k$ pointing up $z$ and $N-k$ pointing down in the $z$-direction, so that the magnetization is $2k-N$. The pointer's state is thus merely its initial state displaced by $k$: $\ket{q=2k-N}=\Phi\big(x-(2k-N)\big)$. 

The usual quantum measurement story goes then on as follows. The pointer being macroscopic, one can directly look at it. If one finds it at position $x_p$, then the state of the measured system collapses to the unnormalised state (its norm square being the probability of finding $x_p$):
\beq \label{Collapse_xp}
\Psi_{S|x_p}=\sum_{k=0}^N \braket{k,N}{\vec m^{\otimes N}} \Phi\big(x_p-(2k-N)\big)\cdot \ket{k,N}
\eeq

A measurement is strong if the pointer has a well defined position, i.e. if its mean square deviation is small with respect to the distance between the eigenvalues of the measured operator: $\Delta << 1$. In this case the sum in (\ref{Collapse_xp}) reduces to a single value $k\approx x_p$, because for all other values of $k$, $\Phi\big(x-(2k-N)\big)$ is (practically) zero. This corresponds to the standard textbook measurement process.

A measurement is weak if, on the contrary, the pointer's position has a quantum uncertainty much larger than the distance between the eigenvalues of the measured operator: $\Delta >> 1$. In this case, many terms in (\ref{Collapse_xp}) remain. Actually, for the most likely results $x_p$ all significant terms remain quasi unchanged. Hence, weak measurements practically don't perturb the $N$-spin system. This is how one can measure the magnetization of magnets.

As a first example, assume $\vec m=\vec e_z$, i.e. all $N$ spins are up in the $z$-direction. Then $\braket{k,N}{\vec m^{\otimes N}}$ vanishes for all $k$ except $k=N$, hence (\ref{SystPointer}) simplifies to:
\beq \label{Collapse_z}
\Psi_{SP}=\ket{\vec e_z^{~\otimes N}}\otimes\ket{q=N}
\eeq
In this case the spin system is not perturbed at all and the pointer moves $N$ steps to the right.

As second example, consider a magnet in the $x$-direction, i.e. $\vec m=\vec e_x$, and a weak measurement with $\Delta\ge\sqrt{N}$. In this case the scalar product $\braket{k,N}{\vec m^{\otimes N}}$almost vanishes except for $k\approx N/2\pm\sqrt{N}$ in which cases $\Phi\big(x_p-(2k-N)\big)\approx\Phi\big(x_p\mp\sqrt{N}\big)$ is essentially independent of $k$. Hence, the first terms in eq. (\ref{SystPointer}) are non-negligible only when the second term is independent of $k$. Consequently, the pointer's central position doesn't move, but merely broadens a bit. Its mean square deviation after the interaction is the convolution of the initial spread $\Delta$ and the square root of the number of spins: $\sqrt{\Delta^2+N}$. Again, the state of the $N$ spins is almost not perturbed.

In summary, weak measurements, as we recalled their formalization, allow one to discriminate magnets pointing to any of the 4 directions $\pm z$ or $\pm x$ without significantly perturbing their quantum state.

\section{weak measurements on $N$ half singlets}\label{WMsinglet}
%===============================================
Let Alice and Bob share $N$ pairs of spins, each in the singlet state. Alice can chose between measuring all her spins individually in the $z$- or in the $x$-directions, i.e measure $\sigma_z$ or $\sigma_x$ on each spin. Adding all her results, on average she should find zero. But in any run (a run consists of $N$ measurement, one on each of her spins), she will find a fluctuation, typically $\pm \sqrt{N}$. Hence, Bob's $N$ spins will result in a magnetization of about $\mu=\pm \sqrt{N}$ in either the $z$- or the $x$-direction, depending on Alice's choice. If Bob could use weak measurements to determine this direction, there would be signalling. How is it that signalling is impossible, despite the fact that Alice's measurement does indeed trigger an arbitrarily large magnetization on Bob's side?

If $\mu$ is the magnetization, i.e. the difference between the number of spins up and down along any direction, then one has $\frac{N+\mu}{2}$ spin up and $\frac{N-\mu}{2}$ spin down along that direction. Assume first that this direction is the $z$-direction. Then, according to the formalism recalled in the previous section \ref{WM}, the pointer will move a distance $\mu$ without broadening and without perturbing the state of the $N$ spins. Hence, the probability distribution of the pointer's position, condition on a magnetization $\mu$, reads:
\beq
\rho^z(x_p|\mu)=|\Phi(x_p-\mu)|^2
\eeq
where the suffix $z$ recalls that Alice measured her spins along the $z$-direction.

Since the probability of a magnetization $\mu$ is binomial: $2^{-N}\binomial{N}{j}$, with $j=\frac{N+\mu}{2}$, Bob's pointer distribution reads:
\beq\label{rhoz}
\rho^z(x_p)=2^{-N}\sum_{j=0}^N\binomial{N}{j}|\Phi\big(x_p-(2j-N)\big)|^2
\eeq

Next, assume that Alice chooses the $x$-direction, hence that the magnetization is along the $x$-direction:
\beq\label{psix}
\Psi_{in}^x=\ket{+x}^{\otimes\frac{N+\mu}{2}}\otimes\ket{-x}^{\otimes\frac{N-\mu}{2}}
\eeq
where\footnote{Note that it is not necessary to symmetrize $\Psi_{in}^x$; indeed, the system-pointer interaction being symmetric, a symmetrized $\Psi_{in}^x$ would lead to the same effect.}
\beq
\ket{\pm x}^{\otimes k}=2^{-k/2}\sum_{j=0}^k \sqrt{\binomial{k}{j}} (\pm1)^{k-j} \ket{j,k}
\eeq
Accordingly, in the $z$-basis  $\Psi_{in}^x$ reads:
\beq
 \Psi_{in}^x=\sum_{j=0}^\frac{N+\mu}{2}\sum_{k=0}^\frac{N-\mu}{2} c_{jk}\ket{j,\frac{N+\mu}{2}}\otimes\ket{k,\frac{N-\mu}{2}}
\eeq
where
\beq
c_{jk}=2^{N/2}\sqrt{\binomial{\frac{N+\mu}{2}}{j}\binomial{\frac{N-\mu}{2}}{k}}(-1)^{\frac{N-\mu}{2}-k}
\eeq
The unitary system-pointer interaction results in:
\beqa
&&U(\Psi_{in}^x\otimes\ket{q=0})=\\
&&\sum_{j,k}c_{jk}\cdot\ket{j,\frac{N+\mu}{2}}\otimes\ket{k,\frac{N-\mu}{2}}\otimes\ket{q=2k+2j-N}\nonumber
\eeqa
Accordingly, the pointer's position probability distribution obtains by tracing out the $N$-spin system reads:
\beq\label{rhoxpmu}
\rho^x(x_p|\mu)=\sum_{j=0}^{j_m}\sum_{k=0}^{k_m}c_{jk}^2 |\Phi\big(x_p-(2k+2j-N)\big)|^2
\eeq
where $j_m=\frac{N+\mu}{2}$, $k_m=\frac{N-\mu}{2}$ and the suffix $x$ recalls that Alice measured her spins along the $x$-direction.

Note that the $(-1)^{\frac{N-\mu}{2}-k}$ sign in the expression of $c_{jk}$ in (\ref{rhoxpmu}) cancels because only the square of $c_{jk}$ appears in $\rho(x_p)$. Furthermore, the double sum in (\ref{rhoxpmu}) can be reduced to a single sum by using the identity $\sum_{j=0}^s\binomial{\frac{N+\mu}{2}}{j}\binomial{\frac{N-\mu}{2}}{s-j}=\binomial{N}{j}$ and the convention $\binomial{k}{j}=0$ for all $j>k$. For this purpose introduce the variable $s=k+j$ and rewrite the double $\sum_{j=0}^{j_m}\sum_{k=0}^{k_m}=\sum_{s=0}^{j_m+k_m}\sum_{j=0}^s$:
\beqa
&&\rho^x(x_p|\mu)= \nonumber\\
&=&2^{-N}\sum_{s=j}^{j_m+k_m}\sum_{j=0}^s\binomial{j_m}{j}\binomial{k_m}{s-j} |\Phi\big(x_p-(2s-N)\big)|^2 \nonumber\\
&=&2^{-N}\sum_{s=0}^N\binomial{N}{s} |\Phi\big(x_p-(2s-N)\big)|^2
\eeqa

Consequently, Bob's pointer position distribution doesn't depend on the magnetization $\mu$ and is rigorously equal to the case Alice measured along the $z$-direction; this holds for all pointer's state $\Phi(x)$, see (\ref{rhoz}). 

This proves that Bob can't get any information about Alice's choice of measurement direction. The reason is that when Alice choses the $z$-direction, Bob's pointer moves without any deformation by a random distance depending on Alice's result, i.e. depending on the magnetization $\mu$. If, on the other hand, Alice chooses the $x$-direction, then the pointer central position doesn't move, but the noise due to the background spins broadens its distribution by precisely the amount required to make it indistinguishable from the case of a $z$-direction measurement. In other words, Bob's magnetization $\mu\approx\sqrt{N}$ doesn't consist of $\sqrt{N}$ spins in the direction corresponding to Alice's measurement, but is smeared in a bath of $N$ random spins with a $\sqrt{N}$ fluctuation in a the direction chosen by Alice. The large bath of random spins in which Bob's magnetization exists hides the information about Alice's direction.

Note that this result is exact for any number $N$ of spins and for any strength of the measurement, i.e. any function $\Phi(x)$, in particular any $\Delta$.

In summary, an arbitrarily large magnetic moment is not necessarily classical in the sense that it might be fundamentally impossible to determine in which direction it points.

\section{Macroscopic limit of isotropic PR-boxes}\label{MacroPR}
%================================================
We just saw that quantum entanglement doesn't allow for signaling, even in the case when it allows one to prepare arbitrarily large magnetic moments at a distance. The inavoidable noise is precisely sufficient to prevent any information transfer, just as in quantum cloning \cite{NGnocloning} and general quantum dynamics \cite{QdynamicsPRL}. This raises the question whether stronger than quantum correlations would lead to signaling when large ensemble are considered.

Let Alice and Bob share $N$ noisy PR-boxes \cite{PRbox}, with isotropic noise \cite{isoPRbox}. Denote the inputs  $x,y\in\{0,1\}$ and outcomes $a,b=\pm1$. Hence, each PR-box has random marginals and correlation
\beq
P(a\cdot b=(-1)^{x\cdot y}|x,y)=V
\eeq
where the "visibility" $V$ is the "pure PR-box weight", $V\in[0,1]$.

As in the case of $N$ singlets, consider the case where Alice measures all her boxes with either the setting $x=0$ or with setting $x=1$ (i.e. she inputs into all her boxes either $x=0$ or $x=1$). In this way she prepares Bob's ensemble of boxes at a distance. If Bob measures all his boxes with the same input $y$ and sums up all his outcomes, he finds a fluctuation of the order $\pm\sqrt{N}$ around zero. If $x\cdot y=0$, then Alice and Bob's fluctuations are likely to be of the same sign; however, if $x\cdot y=1$, then they are likely to be of opposite signs.  So far, this is very similar to the $N$ singlet case. But if the noise is small enough for the correlations to be stronger than quantum, then one may wonder whether signaling is still excluded. 

At this point one would like to define weak-measurements for large ensemble of PR-boxes. Indeed, as emphasized by Rohrlich \cite{Daniel}, any physical box must be such that when large ensembles are considered, then collective measurements of their global ``magnetization'' should be feasible. Unfortunately, at present one doesn't know how to define the analog of weak measurements for ensembles of PR-boxes, a clear weakness of today's concept of PR-boxes. Nevertheless, it makes good sense to assume that in the macroscopic limit of large enough $N$, the following two quantities on Bob's side can both be measured:
\beq
B_y=\sum_{j=1}^N b_{j|y}
\eeq
where $b_{j|y}$ is the outcome of Bob's $j$'th PR-box when it gets the input $y$. 

If the PR-boxes are noise-free, i.e. $V=1$, when Bob could read Alice's input $x$ from $B_0$ and $B_1$. Indeed, if $x=0$, then $B_0=B_1$, while if $x=1$, then $B_0=-B_1$.

But clearly, if the PR-boxes are noisy enough to be realizable with quantum entanglement, i.e. if $V\le\half(1+\sqrt{\frac{1}{2}})\approx 0.85$, then, as we have seen in the previous section \ref{WMsinglet}, the assumption that $B_0$ and $B_1$ are jointly measurable doesn't lead to signaling. Hence the natural question : "How much noise should PR-boxes have to avoid signaling in the macroscopic limit?".

We shall consider the limit of infinitely many PR-boxes and assume that, in this limit, all 4 quantities $A_0,A_1,B_0,B_1$ can be measured simultaneously, where
\beq
A_x=\sum_{j=1}^N a_{j|x}
\eeq
with similar notations for Alice's $a_{j|x}$.

Hence there exist a well defined (i.e. non-negative) probability distribution $P(A_0,A_1,B_0,B_1)$. This implies that the possibility for Bob to measure simultaneously $B_0$ and $B_1$ doesn't lead to signaling. Indeed, the existence of a global probability distribution excludes violation of a Bell inequality, hence guarantees the existence of a local model \cite{Fine}. This also establishes the connection with the concept of macroscopic-locality \cite{Miguel}.

In the limit of many PR-boxes, thanks to the central limit theorem, the probability distribution $P(A_0,A_1,B_0,B_1)$ is Gaussian, with zero mean:
\beqa\label{PGauss}
&&P(A_0,A_1,B_0,B_1)=\\
&&exp\{-(A_0,A_1,B_0,B_1) K^{-1} (A_0,A_1,B_0,B_1)^t\} \nonumber
\eeqa
where the suffix $t$ indicates the transpose and the correlation matrix is defined as follows:
\beq \label{K}
K\equiv\left(%
\begin{array}{cccc}
  \mean{A_0A_0} & \mean{A_0A_1} & \mean{A_0B_0} & \mean{A_0B_1} \\
  \mean{A_1A_0} & \mean{A_1A_1} & \mean{A_1B_0} & \mean{A_1B_1} \\
  \mean{B_0A_0} & \mean{B_0A_1} & \mean{B_0B_0} & \mean{B_0B_1} \\
  \mean{B_1A_0} & \mean{B_1A_1} & \mean{B_1B_0} & \mean{B_1B_1} \\
\end{array}%
\right)
\eeq
with $\mean{A_0A_1}$ the correlation between $A_0$ and $A_1$ and similarly for all entries. $K$ is clearly symmetric.

The first entry is easy to evaluate: $\mean{A_0A_0}=\mean{\sum_{i,j=1}^N a_{i|0}\cdot a_{j|0}}$. If $i=j$ one has $a_{i|0}\cdot a_{j|0}=1$. If $i\ne j$, in the limit of large $N$ the average vanishes. Hence $\mean{A_0A_0}=N$, and similarly for all 4 diagonal terms of $K$.

The second entry $\mean{A_0A_1}$ can't be evaluated without further assumptions. Hence, let's move to the next entry: $\mean{A_0B_0}=\mean{\sum_{i,j=1}^N a_{i|0}\cdot b_{j|0}}$. If $i=j$, $P(a_{i|0}\cdot b_{j|0}=(-1)^{0\cdot0}=+1)=V$, hence $\mean{\sum_{j=1}^N a_{j|0}\cdot b_{j|0}}=N(2V-1)\equiv Nv$. If $i\ne j$, in the limit of large $N$ the average vanishes. Hence $\mean{A_0B_0}=Nv$. Similarly $\mean{A_0B_1}=\mean{A_1B_0}=Nv$ and $\mean{A_1B_1}=-Nv$.

Consequently, the correlation matrix reads:
\beq
K=N\left(%
\begin{array}{cccc}
  1 & s & v & v \\
  s & 1 & v & -v \\
  v & v & 1 & s \\
  v & -v & s & 1 \\
\end{array}%
\right)
\eeq
where we assume $\mean{A_0A_1}=\mean{B_0B_1}$ and define $s\equiv\mean{A_0A_1}/N=\mean{B_0B_1}/N$. Note that one can prove that this symmetry assumption is not necessary to derive our conclusion, though it is a very natural assumption.

Now, the Gaussian probability $P(A_0,A_1,B_0,B_1)$ is non-negative if and only if the correlation matrix $K$ is non-negative. The eigenvalues of $K/N$ are:
\beqa
1&+&\sqrt{2v^2+s^2+2vs}\\
1&-&\sqrt{2v^2+s^2+2vs}\label{e2}\\
1&+&\sqrt{2v^2+s^2-2vs}\\
1&-&\sqrt{2v^2+s^2-2vs}\label{e4}
\eeqa
These must be non negative. Adding (\ref{e2}) and (\ref{e4}) one gets $2\ge 4v^2+2s^2$. Hence, $1-2v^2\ge s^2\ge0$, thus $v^2\le\half$, i.e.
\beq
v\le\sqrt{1/2}
\eeq
which is Tsirelson's bound (recall $V=\frac{1+v}{2}$) \cite{TsirelsonBound}.

Hence, Tsirelson's bound follows from the physical assumption that in the macroscopic limit all measurements are compatible and that this should not lead to signalling. 

\section{Extension to asymmetric noisy PR-boxes}
%===============================================
The result of the previous section can easily be extended to asymmetric non-signalling boxes with arbitrary noise. It suffices to replace (\ref{PGauss}) by:
\beqa\label{PGauss}
&&P(A_0,A_1,B_0,B_1)=\\
&&exp\{-(\bar A_0,\bar A_1,\bar B_0,\bar B_1) \bar K^{-1} (\bar A_0,\bar A_1,\bar B_0,\bar B_1)^t\} \nonumber
\eeqa
where $\bar A_i\equiv A_i-\mean{A_i}$ and $\bar B_i\equiv B_i-\mean{B_i}$ and the correlation matrix $\bar K$ is constructed as (\ref{K}), but using the $\bar A_i$ and $\bar B_j$ instead of the $A_i$ and $B_j$.

The non-negativity of $\bar K$ is then equivalent to the first step in the hierarchy \cite{hierarchy} characterizing quantum correlations. It is known that, in general, this first step is not sufficient to single out quantum correlations, hence - surprisingly - there are stronger than quantum correlations that have a macroscopic non-signalling limit, as emphasized in \cite{AlmostQ}.

\section{Conclusion and open problems}\label{concl}
%=====================================
Large ensembles of small systems should be jointly measurable in some sort of a macroscopic or classical limit. If not, they are not physical \cite{Daniel}. This is true as well for quantum systems as for systems described by any post-quantum theory. In section \ref{WMsinglet} we illustrated this for large ensembles of spin-$\half$ and showed that indeed, the quantum formalism of weak measurement provides the tool to describe collective measurements and how they carefully are just at the border of not violating the no-signaling principle. In the following section we considered noisy PR-boxes, that is hypothetical boxes with stronger than quantum correlations. In the case of isotropic noise and in the limit of infinitely many boxes we found that the assumption that all collective measurements are compatible leads to non-physical signaling whenever the noise is weak enough for the boxes to share correlations stronger than possible according to quantum theory; that is we recovered Tsirelson's bound. This is physically very nice, however one should be able to get to this result without the $N$ to infinity limit. Furthermore, in the case of non-isotropic noise one doesn't recover the quantum boundary (even in the limit $N\rightarrow\infty$), as already emphasized in \cite{AlmostQ}. This is absolutely remarkable and deserves deeper investigation. In particular, there is an urgent need for a model of collective measurements of large-but-finite ensembles of noisy PR-boxes.

\small

\section*{Acknowledgment} This work profited from numerous comments from colleagues from Bristol and Barcelona and discussions after my presentation at the DIQIP meeting in May 2013. I am especially in debt to Sandu Popescu for numerous stimulating discussions and debates. Financial support by the European projects ERA-NET DIQIP and ERC-AG MEC are gratefully acknowledged.

\end{document}